\documentclass[twocolumn,10pt,aps,prl,nofootinbib]{revtex4-1}

\usepackage{amssymb, amsmath, amsthm}
\usepackage{graphicx}
\usepackage{units}

\begin{document}

\title{Maximal refraction and superluminal propagation in a gaseous nanolayer}

\author{J. Keaveney$^{1}$}
\author{I. G. Hughes$^{1}$}
\author{A. Sargsyan$^{2}$}
\author{D. Sarkisyan$^{2}$}
\author{C. S. Adams$^{1}$}
\affiliation{$^{1}$ Joint Quantum Centre (JQC) Durham-Newcastle, Department of Physics, Durham University, South Road, Durham, DH1 3LE, United Kingdom}
\affiliation{$^{2}$ Institute for Physical Research, National Academy of Sciences - Ashtarak 2, 0203, Armenia}

\date{\today}

\begin{abstract}
We present an experimental measurement of the refractive index of high density Rb vapor in a gaseous atomic nanolayer. We use heterodyne interferometry to measure the relative phase shift between two copropagating laser beams as a function of the laser detuning and infer a peak index $n = 1.26 \pm 0.02$, close to the theoretical maximum of 1.31. The large index has a concomitant large index gradient creating a region with steep anomalous dispersion where a sub-nanosecond optical pulse is advanced by  $>100$~ps over a propagation distance of 390~nm, corresponding to a group index $n_{\rm g}=-(1.0\pm0.1)\times10^{5}$, the largest negative group index measured to date.
\end{abstract}

\maketitle



Controlling the speed of light in a medium is key to applications in quantum and optical communication and computation. For optical pulses, the group velocity $v_{\rm g}$ determines the speed of the peak, and depends on the variation of the refractive index $n$ with frequency $\omega$. Around a resonance $v_{\rm g}$ can vary significantly, from slow-light ($v_{\rm g}<c$) \cite{Hau1999}, to fast-light ($v_{\rm g}>c$) \cite{Wang2000}, or even backwards-light ($v_{\rm g}<0$)  \cite{Gehring2006}. 
Employing techniques such as electromagnetically induced transparency \cite{Harris1990,Fleischhauer2005}, where a control laser is used to modify the atomic coherence, 
light has been slowed experimentally to 17 m/s, corresponding to $n_{\rm g}>10^{7}$ \cite{Hau1999}. 
By tuning the control field adiabatically, it is possible to store light in the form of an atomic excitation and retrieve it at a later time \cite{Fleischhauer2000}, forming a quantum memory \cite{Julsgaard2004}.
Much effort has gone into other applications of slow light, such as tunable optical delay lines \cite{Camacho2007} and polarisation control \cite{Siddons2009a}.
%
When $-{\rm d}n/{\rm d}\omega < n/\omega$ the peak of the pulse exits the medium faster than it would have done at light-speed, leading to the term `superluminal'. This surprising phenomenon has led to a wealth of research, including topics such as how fast information can be transmitted in a fast-light medium (signal velocity) \cite{Stenner2003}, and the quantum mechanical implications for noise properties in a slow- or fast-light regime \cite{Kuzmich2001,Boyd2010}. 
%
\begin{figure*}[Ht]
 \includegraphics[width=0.75\textwidth,angle=0]{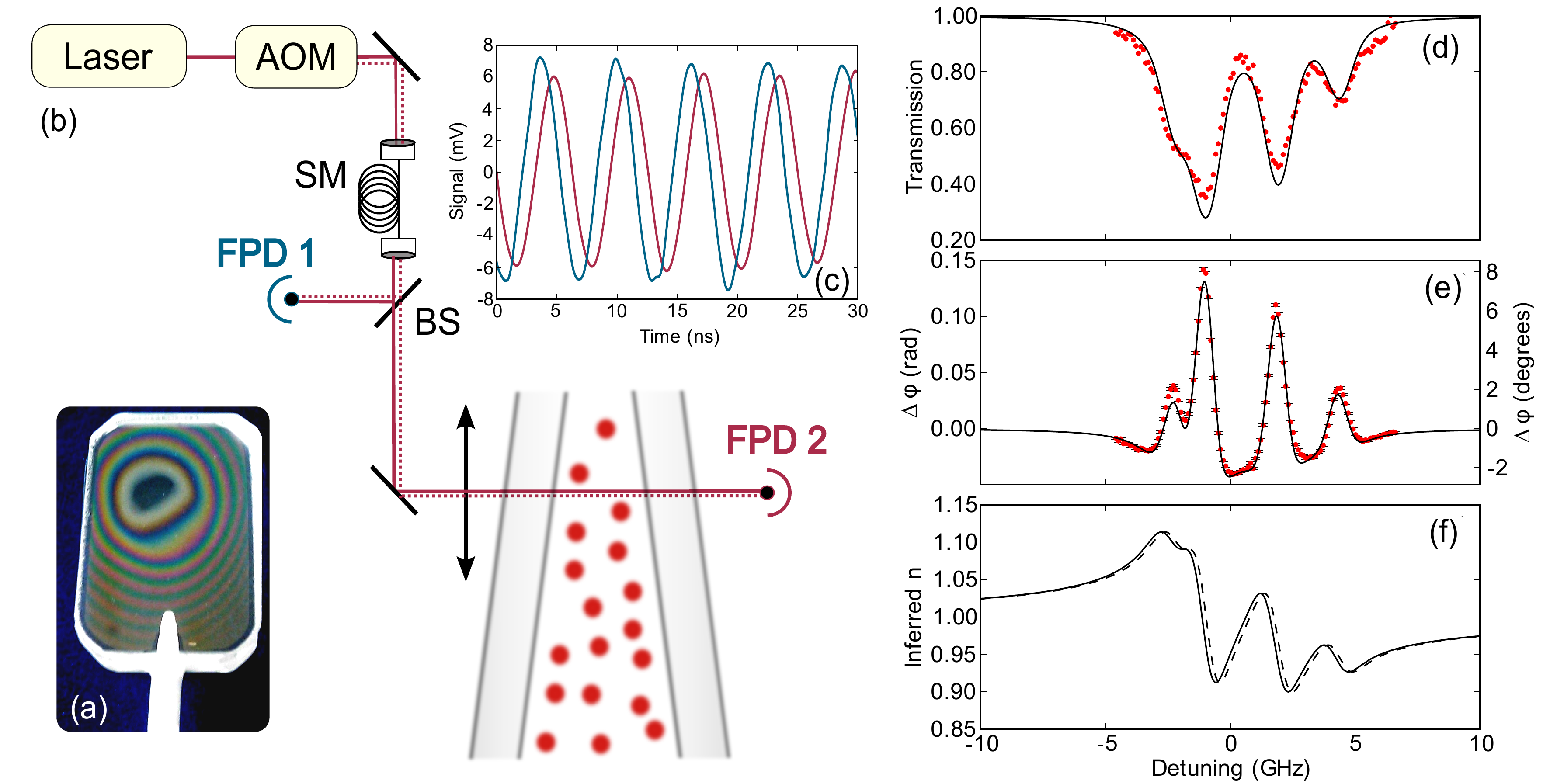}
 \caption{Experimental setup and example data. (a)~Photograph of experimental nanocell used in the experiment. The cell has a wedge-profile (shown schematically) resulting in a tunable vapor thickness. (b)~Schematic of experimental setup. An acousto-optic modulator (AOM) generates a second beam at a detuning $\Delta_{\rm AOM}=160$~MHz which is combined with the unshifted beam through a single mode fiber (SM). After the fiber the beams are mode-matched. A 50:50 beamsplitter (BS) sends half the light to a fast photodiode (FPD1) while the other half propagates through the experimental nanocell. The light is detected after the cell on a second fast photodiode (FPD2). The 160~MHz beat frequency is shown in panel (c) for both beams. We measure the phase shift between the two detectors, and scan the laser frequency over the D2 resonance to obtain transmission (d) and relative phase shift (e) information, shown here for a vapor thickness $\ell=\lambda/2$ and $T=250^{\circ}$C across the D2 resonance, fitted to theory (solid black lines). Panel (f) shows the inferred refractive indices for the shifted (dashed) and unshifted (solid) beams - the solid line in (e) is the difference of these two curves. Zero on the detuning axis represents the weighted line center of the D2 line.}
\label{fig:setup}
\end{figure*}

One way to achieve satisfy the inequality $-{\rm d}n/{\rm d}\omega < n/\omega$
is to center the pulse frequency at the center of an absorption line. However, this 
means that a large gradient ${\rm d}n/{\rm d}\omega$ is accompanied by large absorption.
Achieving a large group index in a medium without large absorption is a key topic, and one which has attracted considerable attention over the past decades. Most solutions to this problem have utilized atomic coherence and quantum interference in multilevel excitation schemes, proposed by Scully \cite{Scully1991a} and first experimentally demonstrated by Zibrov \textit{et. al.}~\cite{Zibrov1996}, creating a region with vanishing absorption with non-zero dispersion on resonance.  
Using a similar concept, the first experiment to measure successfully the superluminal propagation used the anomalous dispersion region in between two narrow gain lines \cite{Wang2000}, where a group index $n_{\rm g}=-310$ was measured. Similarly, backward propagation of pulses was first reported in a gain medium using an erbium-doped fiber amplifier \cite{Gehring2006}, with a measured index of $n_{\rm g}=-4\times 10^{3}$.
All these techniques, however, suffer from the disadvantage that the absolute variation of the refractive index $\Delta n$ is small and as such are limited to low bandwidth $\Delta \omega$, which can be expressed as
\begin{align}
\Delta \omega \approx \omega \dfrac{\Delta n}{n_{\rm g}}~,
\end{align} 
assuming $\vert {\rm d}n/{\rm d}\omega\vert \gg n/\omega$. Thus for a large group index with large bandwidth, a large index enhancement is required. In thermal vapors with a single resonance line, large fractional delays of up to 80 with GHz bandwidth have been demonstrated for slow-light by using off-resonant excitation \cite{Camacho2007}. In principle the maximum observable fractional advance is limited to around 2 by a signal-to-noise argument \cite{Boyd2007}, whilst experimentally fractional advances up to 0.25 have been observed \cite{Tanaka2003}, but with severe attenuation (2\% transmission).

In this work, by using a dense atomic vapor with sub-wavelength thickness we observe an enormous index enhancement $\Delta n= 0.26 \pm 0.02$. As we are in the cooperative regime \cite{Keaveney2012,Pritchard2010}, the dipole-dipole interaction-induced broadening gives rise to a GHz-bandwidth region with very little group velocity dispersion. This allows superluminal propagation of sub-nanosecond pulses with little distortion, and we observe over 100~ps advance across a distance of 390~nm, corresponding to $n_{\rm g}=-(1.0\pm0.1)\times10^{5}$, the largest negative group index measured to date and close to the maximum predicted by a weak probe theoretical model.

\begin{figure}[tb]
 \includegraphics[width=0.49\textwidth,angle=0]{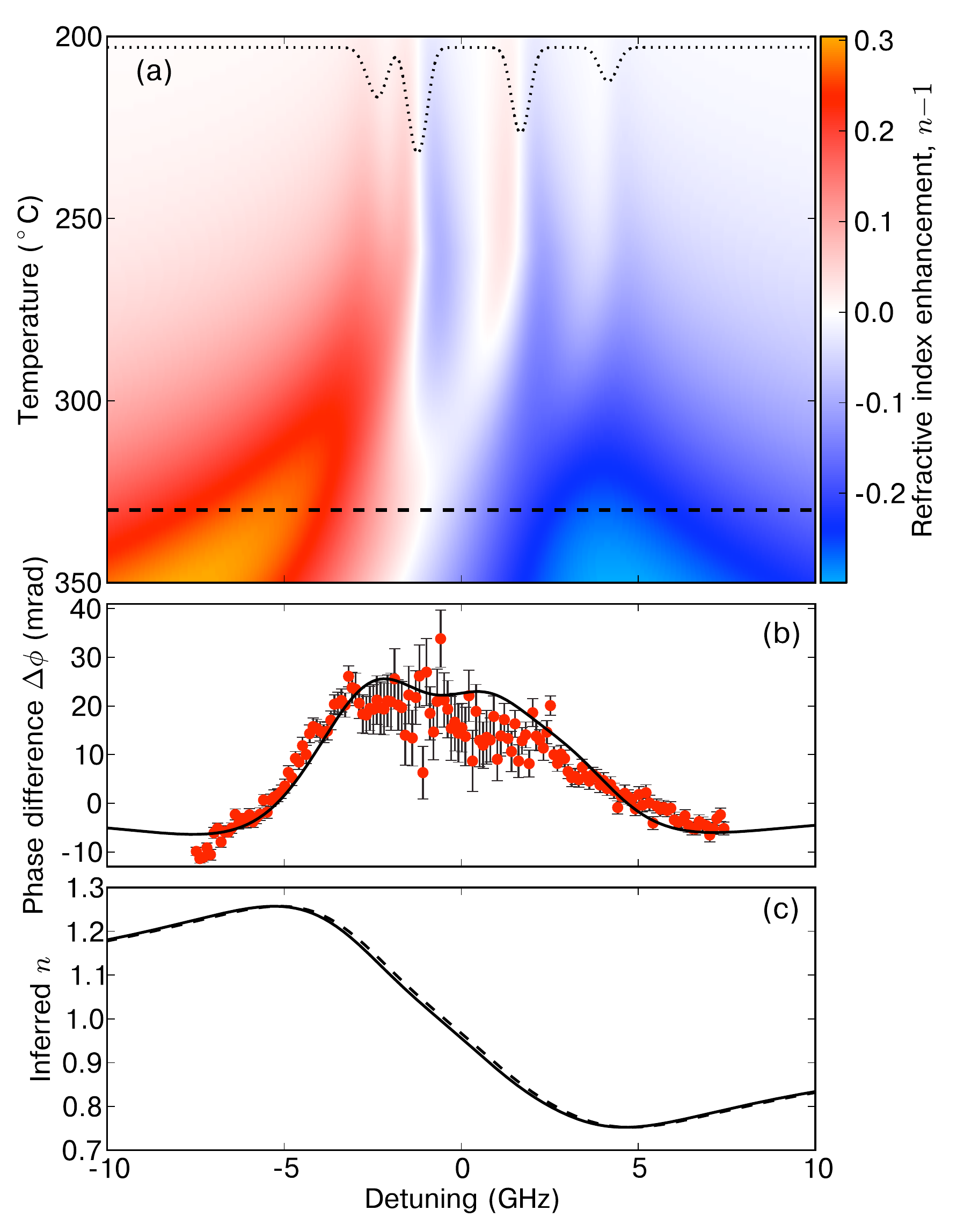}
 \caption{Large refractive index enhancement. (a)~Calculated refractive index as a function of temperature and detuning for the D2 resonance line in Rb. The maximum predicted index is 1.31. The density-dependent redshift can clearly be seen, and has a significant effect on the position of the maximum index. The dotted line shows the positions of the unshifted Doppler broadened resonance lines. (b,~c)~Experimental data for a thickness $\ell=250$~nm and $T=330^{\circ}$C ($N=5\times10^{16}$~cm$^{-3}$, dashed line in (a)). Close to resonance, the large optical depth reduces the signal to the point where accurate phase information is lost. Despite this, the fit to theory is reasonable and we infer a maximum index $n=1.26 \pm 0.02$.}
 \label{fig:maxN}
\end{figure}

Figure \ref{fig:setup} shows the experimental setup. We circumvent the problem of high optical depth by using a very thin layer, much less than the wavelength of the excitation light. 
The atomic vapor layer is confined between two sapphire plates with a tunable separation between 30~nm and 2~$\mu$m (the cell is shown in fig~\ref{fig:setup}a, more details can be found in \cite{Keaveney2012}). By controlling the temperature of the cell, we control the atomic density of the vapor. 
We heat from room temperature to $350^{\circ}$C, corresponding to a number density $N\approx 10^{17}$~cm$^{-3}$.
To measure refractive index, we employ heterodyne interferometry similar to that of Pototschnig \textit{et. al.} \cite{Pototschnig2011}. An acousto-optic modulator (AOM) is used to generate a second beam at $\Delta_{\rm AOM}/(2\pi)=160$~MHz, which is power-matched and then recombined with the probe beam and coupled into a single-mode optical fiber (which ensures mode-matching). Half of the light is then sent to a fast photodiode (FPD1) and the other half focussed to a spot size ($1/{\rm e}^{2}$ radius) of 50~$\mu$m through the cell and onto a second fast photodiode (FPD2). Both detectors measure the beat frequency of the two light fields, and for any one measurement we observe $\sim$300 oscillations. 
The phase of the two signals is then measured as the laser is scanned across the D2 resonance line of Rb ($\lambda = 780.2$~nm). Since the two beams are at nearly the same frequency, the phase difference between the two paths is given by 
\begin{eqnarray}
\phi = \phi_{0} + \Delta\phi = \phi_{0} +\left[n(\Delta+\Delta_{\rm AOM})-n(\Delta)\right]k\ell~, \label{eq:phi}
\end{eqnarray}
where $\phi_{0}$ is an unimportant global phase due to different optical path lengths to the two detectors. Sample data are shown in panel~c. We reconstruct the refractive index profile from fitting the measured relative phase $\Delta \phi$ using equation (\ref{eq:phi}). Panels d,e and f show example data for a vapor thickness equal to half the optical wavelength, $\ell=\lambda/2$. 
For high atomic number densities where the homogeneous broadening is much greater than $\Delta_{\rm AOM}$, we can simultaneously measure the transmitted intensity by measuring the amplitude of the oscillations on FPD2.


%
%

An interesting question that arises when considering an index enhancement this large is what is the maximum possible index of the medium? 
To predict this, we use a model for the susceptibility that includes hyperfine structure, Doppler broadening \cite{Siddons2008b}, Rb-Rb self-broadening \cite{Weller2011a} and magnetic fields \cite{Weller2012}, with excellent agreement to experimental transmission spectra at the 0.5\% level. 
We have also extended this model to include effects of Dicke narrowing, atom-wall interactions and the cooperative Lamb shift  in vapor cells with nanometre thickness \cite{Keaveney2012}.

Whilst other works have speculated that the near-resonance refractive index of a gaseous medium could be ``as high as 10 or 100'' \cite{Fleischhauer1992}, this estimate was based on an independent dipole model and neglected the dipole-dipole interactions which are invariably present at high density. 
Eventually the increase in the susceptibility due to adding more dipoles is exactly cancelled by the higher damping rate due to resonant dipole--dipole interactions \cite{Keaveney2011,OBrien2011a}, so there is no further increase in index. 
A maximum index of $n\approx1.4$ has previously been predicted for Rb \cite{Simmons2012}. 
Using our susceptibility model, we calculate the 
maximum refractive index to be $n=1.31$. This occurs at $T\approx360^{\circ}$C at a detuning $\Delta \sim-7$~GHz from the weighted line center, as shown in fig.~\ref{fig:maxN}(a).
The redshift of the resonance due to the cooperative Lamb shift can also be seen clearly (calculated here for $\ell \gg \lambda$, see \cite{Keaveney2012}) from the figure.
Beyond $T\approx360^{\circ}$C the binary approximation breaks down, and one must consider a multi-perturber model, which is beyond the scope of this work. 
\begin{figure*}[tb]
 \includegraphics[width=0.85\textwidth,angle=0]{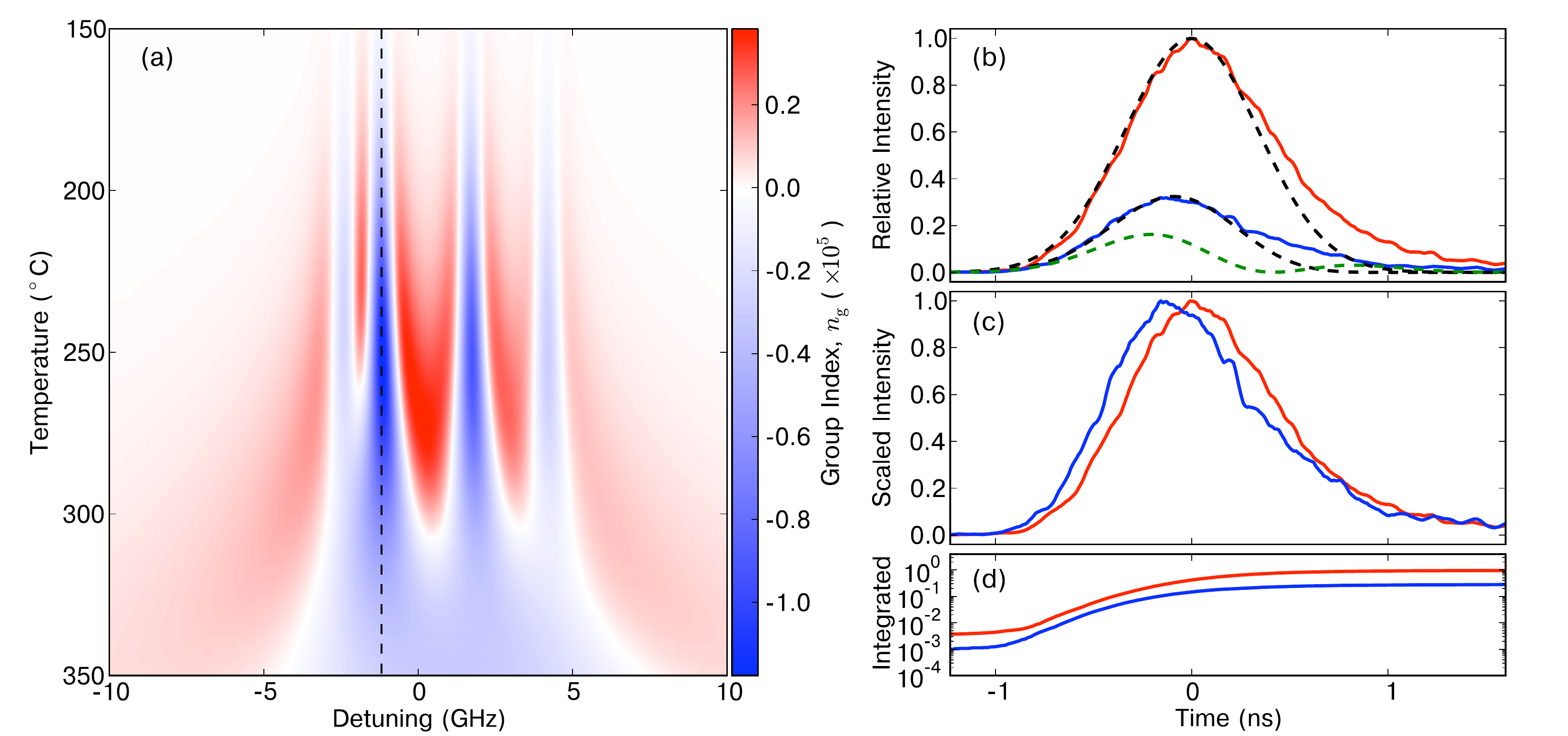}
 \caption{Group index and superluminal pulse propagation. (a) Calculated group index as a function of temperature and detuning. The dashed line represents the carrier detuning for the superluminal pulse. 
 (b,c) Superluminal propagation of a 800~ps pulse through a vapor thickness $\ell=\lambda/2$ and temperature $T=255^{\circ}$C. Red and blue solid lines are experiment, dashed black lines are theory based on a Gaussian input pulse with 800 ps FWHM. Off resonance ($\Delta_{\rm c}=-13$ GHz, red line) there is no interaction at this temperature and the pulse propagates through the vapor as it would through vacuum. On resonance, $\Delta_{\rm c}=-1.2$ GHz, the pulse is attenuated and arrives ($0.13\pm0.01$)~ns earlier than the off-resonance reference pulse, corresponding to a group index $n_{g}=-(1.0 \pm 0.1)  \times 10^{5}$. The dashed green line is the theory without dipole-dipole interactions.
 (d)~Total integrated counts for both signals to verify preservation of causality - the probability of detecting a photon in the advanced pulse is always lower than in the reference pulse, thus the signal velocity is $<c$ \cite{Kuzmich2001}.}
 \label{fig:GI}
\end{figure*}

To observe the maximum index, we require $T=360^{\circ}$C, $N\approx 1 \times 10^{17}$~cm$^{-3}$. Consequently, for these parameters the resonant absorption coefficient is extremely large, and therefore the vapor thickness must be much less than $\lambda$ to transmit a measurable amount of light. 
As one reduces the vapor thickness, however, atom-wall interactions start to have a significant effect on the linewidth, introducing further undesirable broadening and shifts which act to reduce the maximum observable index. 
Despite this, we can still observe a large index, as shown in fig. \ref{fig:maxN}(b,c) for experimental data obtained with vapor thickness $\ell=250$~nm, at $T=330^{\circ}$C. For these conditions, the large on resonance optical depth reduces the signal to the point where a phase cannot be accurately measured. 
Away from resonance, however, we observe good agreement with theory. From this we infer a peak refractive index of $1.26\pm0.02$ approximately 5~GHz red detuned of line center with a transmitted intensity of $\sim40\%$. 
The absorption drops off with detuning much faster than the refractive index decreases. Extrapolating to larger detuning at $\Delta=-15$~GHz we can still have $n\sim 1.13$ with $>90\%$ transmission.
We attribute the difference in theoretical and experimental refractive indices here to additional broadening from the van der Waals atom-surface interaction (included in the theory curves of fig.~\ref{fig:maxN}b/c) which has been characterized for the Rb-Sapphire interface with a model based on a $1/r^{3}$ potential from each cell wall \cite{KeaveneyUNPUB}.

Intuitively, one might anticipate that the largest group index coincides with the largest index enhancement, but this is not the case. From fig.~\ref{fig:GI}a, in which we calculate the group index over the D2 line as a function of temperature and detuning, 
we see a clear maximum around the position of the strongest ground state transition ($^{85}$Rb, $F_{\rm g}=3\rightarrow F_{\rm e}=2,3,4$, $\Delta \approx -1.2$~GHz), where we predict a very large negative group index $n_{\rm g}\approx-1.2 \times 10^{5}$ at $T\approx 255^{\circ}$C. If we tune the laser frequency we can move between fast- and slow-light regimes, but the maximum positive group index is smaller by a factor of $\sim3$.
The spectral broadening at this density creates a region approximately 1~GHz wide with very little group velocity dispersion. While increasing the density further than this increases the bandwidth available, the additional broadening of the line smears out the resonance, reducing the gradient of the index and hence the group index. For clarity, the shift of the resonance due to interactions has been neglected in fig.~\ref{fig:GI}a, since it has a negligible effect on the position of the maximum group index.

We probe this region with a weak optical pulse with a full-width at half-maximum (FWHM) of 800~ps, corresponding to a pulse bandwidth of 1.1~GHz, in a region with vapor thickness $\ell=\lambda/2=390$~nm. 
To measure optical pulses, we lock the laser at an arbitrary frequency electronically using a laser wavelength meter and build up a histogram of the arrival time of photons on a single photon counting module. It is important to note that the bandwidth of the detection equipment is much larger than that of the detected features. We record a reference pulse where the laser is far detuned from resonance so that there is no interaction with the atomic medium. We then switch the laser onto resonance to measure the effect of the medium. 
Figure \ref{fig:GI} panels (b) and (c) show the experimental data. Panel (b) shows both signals normalized to the peak intensity of the reference pulse, while panel (c) shows the pulses scaled to have the same height, to highlight the advance of the resonant pulse. Centering the pulse off resonance (pulse center detuning $\Delta_{\rm c} = -13$~GHz) provides a reference pulse where there is no significant atomic interaction ($n_{\rm g}\approx 1$). On resonance, the peak is advanced, arriving at a time $t_{\rm d}=(-0.13\pm0.01)$~ns earlier than the vacuum pulse. Using $t_{\rm d}=n_{\rm g}\ell/c$, we have a group index $n_{\rm g}=-(1.0 \pm 0.1)\times 10^{5}$.
The dashed lines in fig \ref{fig:maxN}(b) are theory curves based on a Fourier transform method and the susceptibility model, assuming a Gaussian input pulse with 800~ps FWHM, which agree well with the data on the rising edge. On the falling edge, the structure is complicated by fluorescence from the decay of the excited state, which has an exponential decay with a lifetime $\tau\approx1$~ns, corresponding to the time of flight of atoms across the cell.
The large spectral bandwidth available means we see very little distortion of the resonant pulse. 
It is important to note that without the dipole-dipole interactions, the propagation of such temporally short pulses would not be possible without heavy distortion of the pulse shape. 
To illustrate this, the green dashed line in fig.~\ref{fig:GI}b shows the calculated pulse profile if there were no dipole-dipole interactions in the medium. In this case the pulse has a much larger bandwidth than the transition (and spans multiple excited state hyperfine levels) and therefore distortion of the output pulse is evident. Even though there is a superluminal component which is more advanced than for the interacting ensemble, there is also a sub-luminal component which can be seen as a small peak at $t\sim 0.8$~ns.
Panel~d shows the total integrated counts over the detection period, which verify that the advanced pulse preserves causality \cite{Kuzmich2001}, since there is always a greater probability of detecting a photon in the reference pulse than the advanced pulse. From this we can immediately surmise that the information velocity of the advanced pulse is less than $c$.

%

%
%

In demonstrating this giant refractive index and group index, we open the door to further investigation of slow- and fast-light effects in the cooperative interaction regime, where dipole-dipole interactions play an important role \cite{Keaveney2012}. 
The GHz bandwidth available enables control of the dynamics on a faster timescale than the natural lifetime of the excited state (27~ns). Since the cooperative effects are dependent on the level of excitation in the medium \cite{Friedberg1973}, a pump-probe setup with a strong and weak pulse with a short time delay may yield interesting physics, such as sub- or superluminal gain in a transiently inverted medium. 
This will form the basis of future research.

We would like to thank V. Sandoghdar for stimulating discussion and acknowledge financial support from EPSRC and Durham University.






%

\end{document}